\documentclass[1, 12pt]{elsarticle}

\usepackage{hyperref}

\usepackage{amssymb}
\usepackage{amsmath}
\usepackage[utf8]{inputenc}

\biboptions{comma,square,sort&compress}

\journal{Progress in Materials Science, accepted for publication}

\begin{document}

\begin{frontmatter}

\title{Phase-field modeling of microstructure evolution: \\ Recent applications, perspectives and challenges}

\author{Damien Tourret$^{a, 1\fnref{CE}}$}
\author{Hong Liu$^{b, 1 \fnref{CE}}$}
\author{Javier LLorca$^{a, c, }$\corref{cor1}}
\address{$^a$ IMDEA Materials Institute, C/ Eric Kandel 2, 28906, Getafe, Madrid, Spain. \\  
$^b$ National Engineering Research Center of Light Alloy Net Forming and State Key Laboratory of Metal Matrix Composite, Shanghai Jiao Tong University, 200240 Shanghai, PR China \\
$^c$ Department of Materials Science, Polytechnic University of Madrid, E. T. S. de Ingenieros de Caminos. 28040 - Madrid, Spain.}

\cortext[cor1]{Corresponding author}
\fntext[CE]{Authors contributed equally to the paper.}

\begin{abstract}

We briefly review the state-of-the-art in phase-field modeling of microstructure evolution. The focus is placed on recent applications of phase-field simulations of solid-state microstructure evolution and solidification that have been compared and/or validated with experiments. They show the potential of phase-field modeling to make quantitative predictions of the link between processing and microstructure. Finally, some current challenges in extending the application of phase-field models within  the context of integrated computational materials engineering are mentioned.

\end{abstract}


\end{frontmatter}

\section{Introduction}

The processing-microstructure-properties linking is the central paradigm of Materials Science, indicating that the arrangement and properties of the different phases at the microscopic level determines to large extent the macroscopic behavior of materials. The main building block of most materials are crystalline regions whose properties are determined by the presence of defects in the lattice (such as solute atoms, dislocations, secondary phases, precipitates, etc.) as well as by the lattice boundaries (dendrites, grain and twin boundaries, etc.). It was understood early that numerical simulations at the ``mesoscale'' (encompassing the interaction of the homogeneous crystalline lattice with defects) were necessary to understand the complex processing-structure-properties relationships as well as to guide the design of materials with outstanding properties. Most of the effort in linking microstructure to properties has relied on the finite element method (and similar approaches, such as using fast Fourier transforms) in which the mesoscopic domain is divided in homogeneous regions limited by boundaries and the response of each domain is controlled by the appropriate partial differential equations. The overall response of the material is determined by solving the boundary value problem in which the solution to the differential equations also satisfies the boundary conditions \cite{NH99, SLL18}. This strategy has been extremely successful for analyzing the microstructure-properties links as long as the topology of the boundaries between different homogeneous regions do not change. However, this is usually not the case during processing as the evolution of the boundaries is precisely part of the solution of the problem.

Since the early 90's, the phase-field (PF) method has emerged as an outstanding tool for simulating the formation and evolution of microstructures during processing \cite{chen1996continuum, chen2002phase, moelans2008introduction, qin2010phase, ode2001recent, boettinger2002phase, karma2003phase, singer2008phase, steinbach2009phase, steinbach2013phase, plapp2016phase, provatas2011phase}. 
The key idea of the PF approach is the description of interfaces using continuous fields. Discontinuities of properties across interfaces as well as specific boundary conditions at the interface (found in models that consider sharp interfaces) are smeared out and represented by a smooth variation of one or several auxiliary fields (the phase fields) across a diffuse interface. In this way, the problem is solved by integrating a unique set of partial differential equations in the whole domain, and the evolution of the different interfaces comes out naturally as part of the solution. Hence, the PF method provides a powerful tool to handle notoriously challenging free-boundary problems for arbitrarily complex interfaces. Moreover, its generality and flexibility make the PF approach convenient to apply to a broad range of phenomena across materials science, and beyond, including fracture \cite{chen2017instability, lubomirsky2018universality, mesgarnejad2020vulnerable, emdadi2021phase}, liquid metal dealloying \cite{geslin2015topology, mccue2016kinetics}, sintering \cite{abdeljawad2019sintering, yang20193d, yang2020investigation}, and many other applications \cite{tonks2019phase}. 

In this short and far-from-exhaustive review, we first highlight a few recent applications of the PF method to phase transformations and microstructural evolution, which demonstrate its current potential to make quantitative predictions of the link between processing and microstructure. Particular emphasis is given to microstructure evolution in metals, alloys, and relevant model systems subjected to either solidification or solid-state transformations. In the second part, we briefly review some of the current challenges to extend the application of PF within the framework of integrated computational materials engineering.

\section{Phase-Field Modeling Fundamentals}

The cornerstone of nearly every PF model is the formulation of the free energy $F$ as a functional 
 \begin{align}
F=\int{
f(
\phi_1, \phi_2, \dots, \phi_n,
c_1, c_2, \dots, c_n,
\nabla\phi_1, \nabla\phi_2, \dots, \nabla\phi_n,
\nabla c_1, \nabla c_2, \dots, \nabla c_n,
p, T, \dots
)} \mathop{dV}
\end{align}

\noindent containing a set of non-conserved fields $\phi_i$ (for instance, order parameters describing crystalline symmetries) and conserved fields $c_i$ (for instance, species concentration). The energy density $f$ typically includes (1) a potential with local minima in the two (or more) coexisting phases (for instance ``double-well’’ or ``double-obstacle’’ potentials \cite{steinbach2009phase}), (2) gradient terms of $\nabla\phi_i$ and $\nabla c_i$ that relate to the energetic cost of interfaces and (3) bulk energy density terms as a function of the local fields $\phi_i$ and $c_i$, local state variables (e.g. temperature $T$, pressure $p$, stress/strain), and/or external stimuli. 
The evolution of conserved fields, $c_i$, and non-conserved fields, $\phi_i$, are thus given by a set of equations that guarantees a decrease in total free energy with time according to
\begin{align}
& \frac{\partial c_i}{\partial t} 
= \nabla \cdot \bigg( M_{c_i} \nabla \frac{\delta F}{\delta c_i} \bigg)
= \nabla \cdot \bigg( M_{c_i} \nabla \left[ \frac{\partial f}{\partial c_i} - \nabla \cdot \frac{\partial f}{\partial \nabla c_i} \right] \bigg)
\\
& \frac{\partial \phi_i}{\partial t} 
= - M_{\phi_i} \frac{\delta F}{\delta \phi_i}
= - M_{\phi_i} \left[ \frac{\partial f}{\partial \phi_i} - \nabla \cdot \frac{\partial f}{\partial \nabla \phi_i} \right]
\end{align}
respectively referred to as Cahn-Hilliard \cite{cahn1958free, cahn1959free, cahn1961spinodal} and Allen-Cahn \cite{cahn1977microscopic, allen1979microscopic} (or time-dependent Ginzburg-Landau) equations. Parameters $M_{c_i}$ and $M_{\phi_i}$ are positive mobilities, e.g. linked to atomic ($c_i$) or interfacial ($\phi_i$) mobilities, and may depend on local state variables and interface orientation. 

The main differences between a given phase-field model and another reside primarily in the set of conserved and non-conserved fields chosen to describe the problem at hand, the resulting formulation of the various contributions to the total free energy, and the choice of interpolation of the different variables across interfaces. Microstructural evolution is driven by a reduction of the system free energy --- namely a reduction of bulk free energy for phase transformations, a reduction of interfacial energy for coarsening or grain growth, and often a subtle combination of both. The general form of the energy functional is a decisive advantage of the method, permitting to couple a broad range of distinct phenomena in a consistent framework. 

\section{Phase-Field modeling of Solid-State Microstructure Evolution}

\subsection{Fundamental concepts}

Phase-field models for solid-state microstructural evolution commonly derive from the microscopic elasticity theory of structural transformations by Khachaturyan \cite{khachaturyan1968microscopic, khachaturyan1983theory, chen1991computer, wang1993kinetics}. Such models involve actual order parameters, $\phi_i$, representing long-range atomic ordering as well as symmetry and orientation relationships between coexisting phases, which naturally arise for many solid-solid transformations. Different transformations may be represented by a different number of non-conserved order parameters, $\phi_i$, which may be coupled to conserved parameters, $c_i$ \cite{chen1996continuum, chen2002phase, moelans2008introduction}. 

The fields $\phi_i$ thus provide a valuable description of the crystalline structure of the different phases, and the anisotropy of interfacial properties (e.g. free energy) derives naturally from the gradient terms in the free energy. On the other hand, beyond a few relatively simple cases (e.g. anti-phase boundaries), the energetics of phase transformations need to be described phenomenologically, thus introducing unknown parameters that may require identification through dedicated experiments or theories. 
In most models, conserved and non-conserved fields are coupled in a way that prevents changing the interface width and its energy independently, hence limiting the upscaling of the diffuse interface width for computational purposes \cite{moelans2008introduction, wheeler1993phase}.
Previous detailed reviews of the PF method oriented toward solid-state microstructure evolution can be found in \cite{chen1996continuum, chen2002phase, moelans2008introduction}.

\subsection{Highlighted applications}

The phase-field method has been applied to a broad range of solid-state microstructural evolution phenomena. In this review, we highlight recent studies of four mechanisms, including a diffusive transformation (precipitation), a displacive transformation (martensitic), a coarsening mechanism (grain growth), and a deformation mechanism (twinning). 
Other notable applications include, for instance, the modeling of dislocations and their interaction with microstructure evolution \cite{hu2001solute, wang2001phase, rodney2003phase, wang2010phase, beyerlein2016understanding, hunter2018review} and the introduction of  plastic and viscoplastic effects \cite{guo2005elastoplastic, zhou2008contributions, ammar2009combining, gaubert2010coupling, ammar2011phase, shanthraj2016phase}. 

\subsubsection{Precipitation}

Precipitation is a solid-state diffusive phase transformation used in a broad range of metallic alloys due to the strengthening effect of precipitates. 
If precipitates form homogeneously in the matrix, they have a random spatial distribution. 
However, interactions among nearby precipitates and/or pre-existing lattice defects, such as dislocations and grain boundaries, may result in non-random distributions. 
For example, the transformation strain from precipitation may result in an increase of nucleation energy barrier adjacent to the precipitate. 
The precipitate may interact with neighboring ones and/or form on pre-existing lattice defects to decrease this barrier. 

The phase-field method has been used extensively to simulate precipitation in a wide range of metallic alloys, such as Ni \cite{li1997shape, wang1998field, zhu2004three, wang2008coarsening, fromm2012linking, wen2006phase, zhou2008contributions, vorontsov2010shearing, zhou2011modeling, zhou2007phase, zhou2010large, boussinot2009phase, cottura2012phase, cottura2016coupling, degeiter2020instabilities, boisse2007phase, ali2020role, ali202045, mohanty2008phase, kitashima2009new, vorontsov2012shearing, kundin2012phase}, 
Ti \cite{shi2013variant, boyne2014pseudospinodal, shi2015microstructure, qiu2015variant, qiu2016effect, shi2019growth, teixeira2006modeling, teixeira2007transformation}, 
Al \cite{li1998computer, vaithyanathan2002multiscale, vaithyanathan2004multiscale, ji2018phase, liu2017multiscale, RV18, EMP19, kim2017first, liu2015phase, liu2019precipitation}, 
and Mg \cite{gao2012simulation, ji2014predicting, liu2013simulation, liu2017structure, liu2017formation, dewitt2017misfit, liu2015guided,  liu2017simulation, han2013phase, han2015three, han2015large} alloys. 
 Research studies have covered a series of aspects related to the growth and coarsening of precipitates, such as the shape and distribution of precipitates \cite{li1997shape,liu2013simulation,liu2017multiscale}, the interactions between precipitations and external stress fields, such as the stress fields caused by other precipitates \cite{liu2017structure, han2013phase} or lattice defects \cite{liu2013simulation,liu2015guided}, and the directional coarsening (rafting) and stability of cuboidal matrix/precipitate microstructures in superalloys \cite{boisse2007phase, boussinot2009phase, zhou2007phase, zhou2010large, cottura2012phase, cottura2016coupling, degeiter2020instabilities, ali2020role, ali202045}.
Here we illustrate the application of the PF method to study the formation of $\theta'$ (Al$_2$Cu) precipitates in Al-Cu alloys. 

\begin{figure}[h]
\centering
\includegraphics[width=4in]{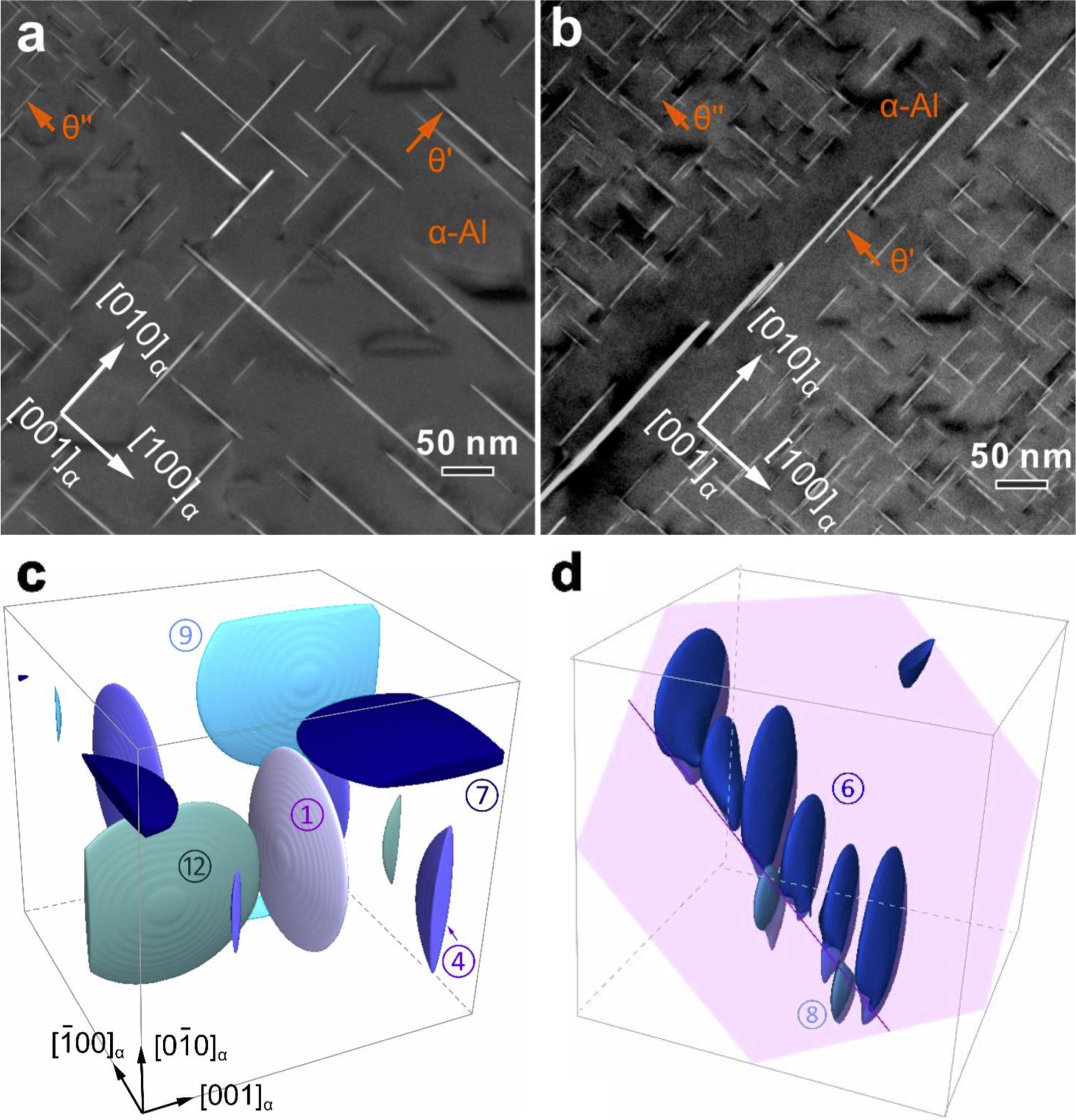}
\caption{Precipitation of $\theta'$ in an Al-Cu alloy \cite{liu2017multiscale}: TEM images showing (a) random and (b) orientation-correlated $\theta'$ precipitates in an Al-4 wt.\% Cu alloy aged at 180$^\circ$C for 66 hours, and phase-field simulation of (c) the homogeneous nucleation of $\theta'$ precipitates, leading to a random precipitate distribution, and (d) precipitates formed under the stress field of mixed dislocations, leading to correlated orientations.} \label{fig:precip} 
\end{figure}

The $\theta'$ is a key strengthening phase in Al-Cu alloys widely used in automobile and aerospace industries over the past century \cite{polmear2017light}. 
To accurately predict the shape and formation mechanism of $\theta'$ precipitates, input parameters are needed that include transformation strains, elastic constants, free energy curves of both $\theta'$ precipitates and $\alpha$-Al matrix, and interfacial energy between $\theta'$ precipitates and $\alpha$-Al matrix.
These parameters can be calculated by atomistic approaches, such as first principles and molecular dynamics/statics, and/or collected from available databases (see e.g. \cite{li1998computer, vaithyanathan2004multiscale, kim2017first, liu2017multiscale, RV18, EMP19}).  
Simulation results have revealed that both the anisotropic interfacial energy and elastic strain energy favor the precipitate growth along the same habit plane, and that shear strain plays a dominant role on the equilibrium shape of $\theta'$ precipitates (see, e.g., Fig.~\ref{fig:precip} \cite{liu2017multiscale}).
It was also shown that pre-existing lattice defects, such as dislocations, can facilitate the nucleation and growth of precipitates, in Al \cite{liu2015phase, liu2017multiscale} (Fig.~\ref{fig:precip}) but also in other alloys \cite{gao2012simulation, liu2015guided, shi2015microstructure, qiu2015variant, li1997shape, zhou2008contributions}.

\subsubsection{Martensitic transformation}

Diffusionless (or displacive) phase transformations occur when a change in crystal structure is induced by a cooperative movement of large groups of atoms.
A common, and likely the most studied, displacive transformation is the martensitic transformation, in which the driving forces can come from either Gibbs free energy differences between the matrix and the secondary phases, or from the external applied stress. 
Due to lattice changes from the matrix to the product phase, this transformation is accompanied by the generation of transformation strain, which is a hindrance for transformation. 

To date, Landau theory and phase-field models have been extensively used to model for martensitic transformations \cite{barsch1984twin, saxena1997hierarchical, shenoy1999martensitic, ahluwalia2003elastic, lookman2003ferroelastic, ahluwalia2004landau, wang1997three, artemev2001three, artemev2002three, wang2004effects, fan1995computer, wen1999effect, gao2012p, gao2014pattern, zhu2017taming, wang2006multi, zhao2017effect, xu2016landau, zhang2007modelling, zhu2017crystallographic, levitas2002three, levitas2011surface, cui2007simulation, shchyglo2012martensitic, shchyglo2019phase, yamanaka2008elastoplastic, yamanaka2010elastoplastic, yeddu2012three, malik2012three, mamivand2013review}. 
Simulations have allowed identifying important mechanisms for martensitic microstructure formation, for instance focusing on the shape and distribution of phases in martensitic microstructures \cite{wang1997three, artemev2001three, artemev2002three, wang2004effects, fan1995computer, wen1999effect, gao2012p, gao2014pattern}, nucleation mechanisms \cite{wang2006multi, zhang2007modelling, zhao2017effect, xu2016landau}, or the effect of plasticity \cite{yamanaka2008elastoplastic, yamanaka2010elastoplastic, yeddu2012three, malik2012three} on transformation kinetics and resulting microstructures.

\begin{figure}[!h]
\centering
\includegraphics[width=4.5in]{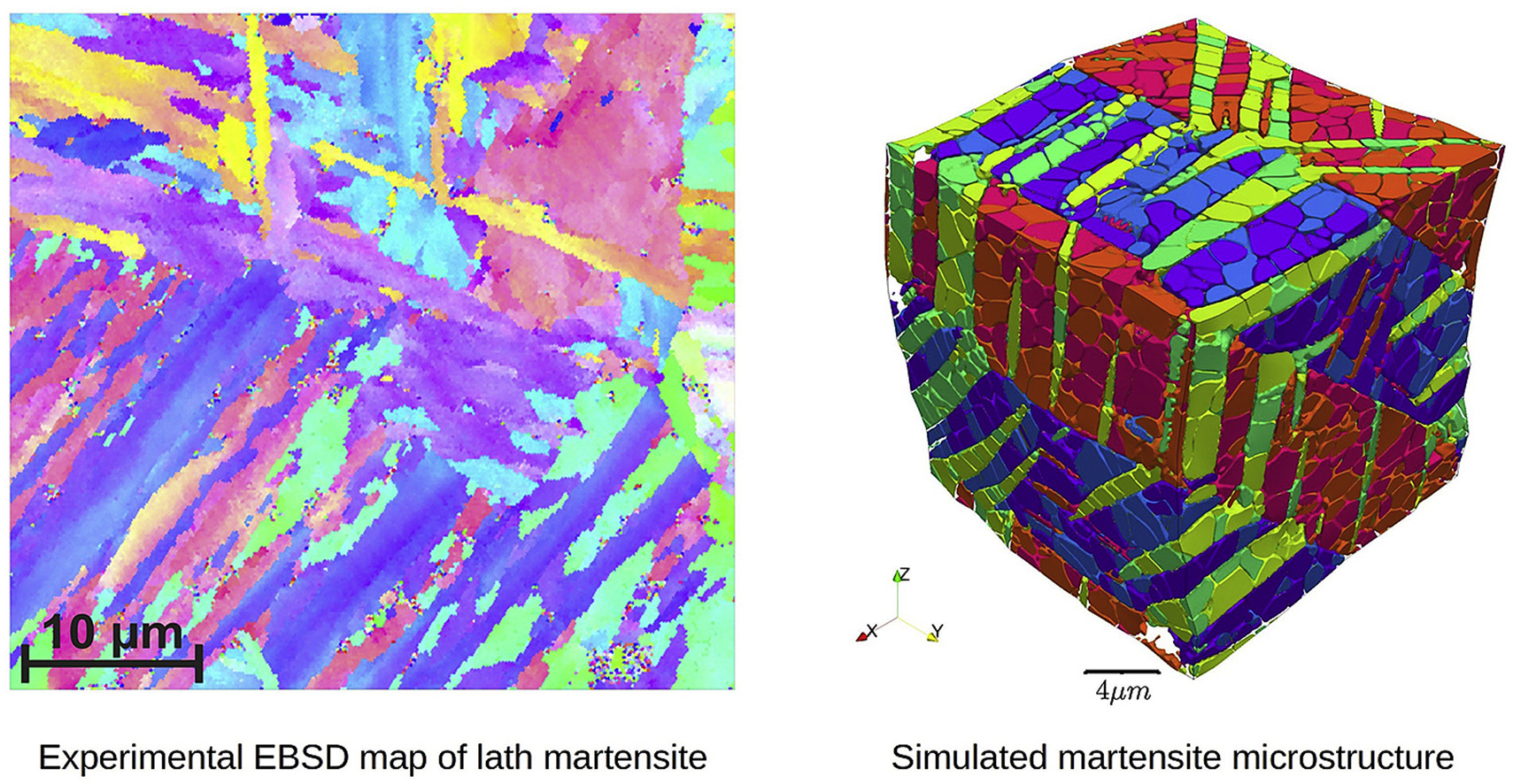}
\caption{Martensitic microstructures in low-carbon steels \cite{shchyglo2019phase}: (left) EBSD orientation maps of a steel containing 0.3 wt\%C and (right) phase-field simulated microstructure with color map showing the 24 Kurdjumov-Sachs orientation variants.
Reproduced with permission from \cite{shchyglo2019phase}}. \label{fig:martensite}.  
\end{figure}

From an industrial perspective, the martensitic transformation in steels is of the utmost importance, as it confers martensitic steels higher strength and hardness than most other steel grades. 
Yet, rapid transformation kinetics makes it challenging to observe {\it in situ}, hence making simulations all the more useful. 
A recent application of the PF method to the martensitic transformation is illustrated in Fig.~\ref{fig:martensite}, where the experimental microstructures observed in low-carbon steels and the simulated microstructures accounting for 24 orientation variants in a finite strain framework are compared \cite{shchyglo2019phase}. 
Beyond steels, notable recent PF studies on martensitic transformations have focused on Ni-Ti-based shape memory alloys \cite{shchyglo2012martensitic, gao2012p, wang2019phase} and tetragonal-to-monoclinic transformation in Zr alloys \cite{mamivand2013phase, mamivand2014phase, cisse2020aphasefield}.

\subsubsection{Grain Growth}

Most technological materials are polycrystalline. Their grain size is critical to determine their properties, and it can be modified through annealing (i.e heating), which leads to the coarsening of the microstructure \cite{atkinson1988overview, humphreys2012recrystallization}. 
This phenomenon is referred to as grain growth, and its main underlying mechanism is clear: the reduction in free energy through a decrease of grain boundary (GB) area. 
GB energies are known to depend on the misorientation of the grains and on the inclination of the GB plane \cite{rohrer2011grain, ratanaphan2015grain}. Their mobilities are also assumed to depend mostly on the five degrees of freedom representing the GB misorientation and inclination \cite{rollett2004grain, gottstein2001grain, gottstein2009grain}. PF models provide a compelling tool to investigate grain growth, as they allow the integration of anisotropic GB properties, and their dependence upon GB misorientation and inclination.

A broad variety of PF models have been developed to simulate grain growth 
\cite{chen1994computer, fan1997computer, fan1997computer2, krill2002computer, steinbach1996phase, kobayashi1998vector, lusk1999phase, kim2006computer, suwa2008parallel, kamachali20123, kamachali2015geometrical, yadav2018investigation, yadav2018analysis, miyoshi2017ultra, chen1996computer, fan1997diffusion, fan2002phase, poulsen2013three, yadav2016effect, ravash2014three, ravash2017three, kazaryan2000generalized, kazaryan2002grain, suwa2007three, moelans2008quantitative, zaeem2011investigating, kim2014phase, miyoshi2016validation, miyoshi2016extended, salama2020role, fan1998numerical, chang2009effect, moelans2005phase, moelans2006phase, moelans2007pinning, vanherpe2010pinning, chang2015phase, suwa2006phase, schwarze2016phase, suwa2007phase, ko2009abnormal, liu2019phase, takaki2007phase, takaki2009multi, takaki2014multiscale, moelans2013phase, abrivard2012phase1, abrivard2012phase2}. 
They were used to study coarsening mechanisms in single-phase \cite{kim2006computer, suwa2008parallel, kamachali20123, kamachali2015geometrical, yadav2018investigation, yadav2018analysis, miyoshi2017ultra, chen1994computer, fan1997computer, fan1997computer2, krill2002computer, steinbach1996phase, kobayashi1998vector, lusk1999phase}, two-phase \cite{chen1996computer, fan1997diffusion, fan2002phase, poulsen2013three, yadav2016effect}, and three-phase microstructures \cite{ravash2014three,ravash2017three}, and to explore a variety of phenomena, including the evolution toward a characteristic grain size distribution \cite{kim2006computer, suwa2008parallel, kamachali20123, kamachali2015geometrical, yadav2018investigation, yadav2018analysis, miyoshi2017ultra}, 
the effect of anisotropic GB properties \cite{kazaryan2000generalized, kazaryan2002grain, suwa2007three, moelans2008quantitative, zaeem2011investigating, kim2014phase, miyoshi2016validation, miyoshi2016extended, salama2020role}, 
the effect secondary particles, e.g. Zener pinning \cite{fan1998numerical, chang2009effect, moelans2005phase, moelans2006phase, moelans2007pinning, vanherpe2010pinning, chang2015phase, suwa2006phase, schwarze2016phase}, 
abnormal grain growth \cite{suwa2007phase, ko2009abnormal, liu2019phase}, 
and recrystallization \cite{takaki2007phase, takaki2009multi, takaki2014multiscale, moelans2013phase, abrivard2012phase1, abrivard2012phase2}, 
among other applications. 

\begin{figure}[!h]
\centering
\includegraphics[width=4.5in]{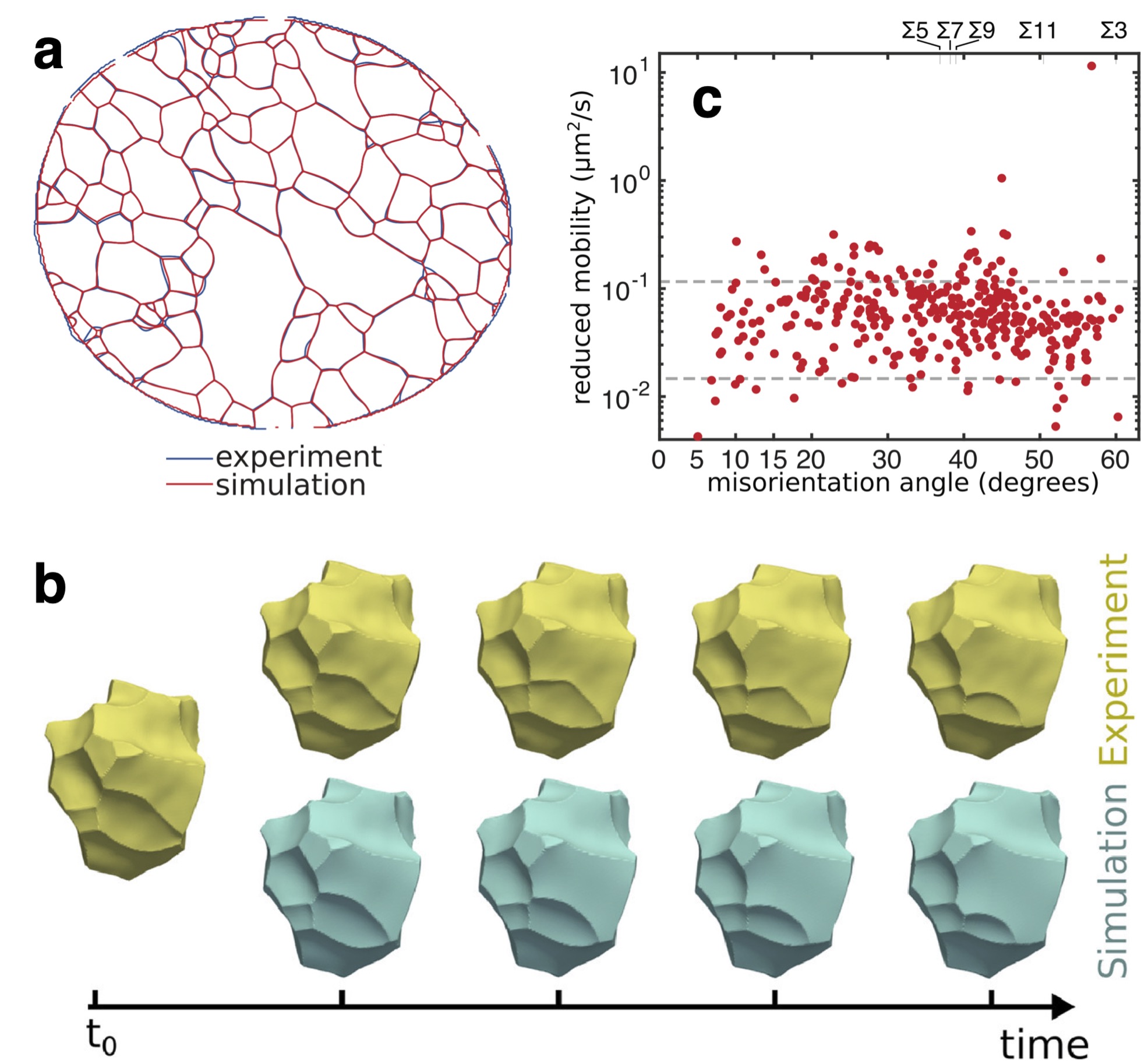}
\caption{
\label{fig:graingrowth} 
Extracting grain boundary mobilities in polycrystals by combining 3D time-resolved tomography and phase-field modeling \cite{zhang20202grain}: (a) experimental and PF simulated grain boundaries within one slice of the 3D sample with optimized/fitted reduced mobilities, (b) morphological evolution of a given grain from tomography (top) and PF simulations with optimized reduced mobilities (bottom), and (c) resulting reduced mobilities of hundreds of grain boundaries as a function of their misorientation angle.
Reproduced with permission from \cite{zhang20202grain}. 
}
\end{figure}

 A recent PF study of grain growth \cite{zhang20202grain} is illustrated in Figure \ref{fig:graingrowth}. Simulations are compared with time-resolved X-ray 3D tomography data recorded during annealing of an iron sample. By fitting simulations and experiments (Fig. \ref{fig:graingrowth}a,b) \cite{zhang20202grain,zhang2017determining}, the authors extract the mobilities of several hundreds of grain boundaries. Remarkably, reduced GB mobilities do not show any correlation with the GB macroscopic degrees of freedom (see, e.g., misorientation in Fig. \ref{fig:graingrowth}c) and some even seem to be time-dependent. These results, which appear to agree with results from atomistic simulations \cite{janssens2006computing, olmsted2009survey}, challenge the conventional assumption that GB mobilities are primarily governed by the GB misorientation and inclination

\subsubsection{Deformation twinning}

Twinning is one of the major deformation modes in metallic materials \cite{christian1995deformation, lebensohn1993study}, in particular for hcp crystals or in nanocrystalline metals \cite{chen2003deformation, yamakov2004deformation, wu2008new, zhu2009formation, nie2013periodic}. The formability, yield strength, and tension-compression yield asymmetry of some metallic materials, such as wrought Mg alloys, are all closely related with deformation twinning. 

In a PF framework, twin crystals are essentially assimilated to martensite plates and the applied stress/strain provides the driving force. The evolution of twin crystals is thus equivalent to the gliding of twinning dislocations. 
The PF method has been applied to study twinning and detwinning in both cubic \cite{hu2010simulations, heo2011phase, gu2013phase, heo2014spinodal} and hexagonal materials \cite{clayton2011phase, clayton2016phase, pi2016phase, liu2018formation}, including under finite strain, e.g. coupling with crystal plasticity theory (see, e.g. \cite{kondo2014phase, liu2018integrated}) or extending Khachaturyan’s classical theory (e.g. \cite{zhao2020finite}).

\begin{figure}[!h]
\centering
\includegraphics[width=\columnwidth]{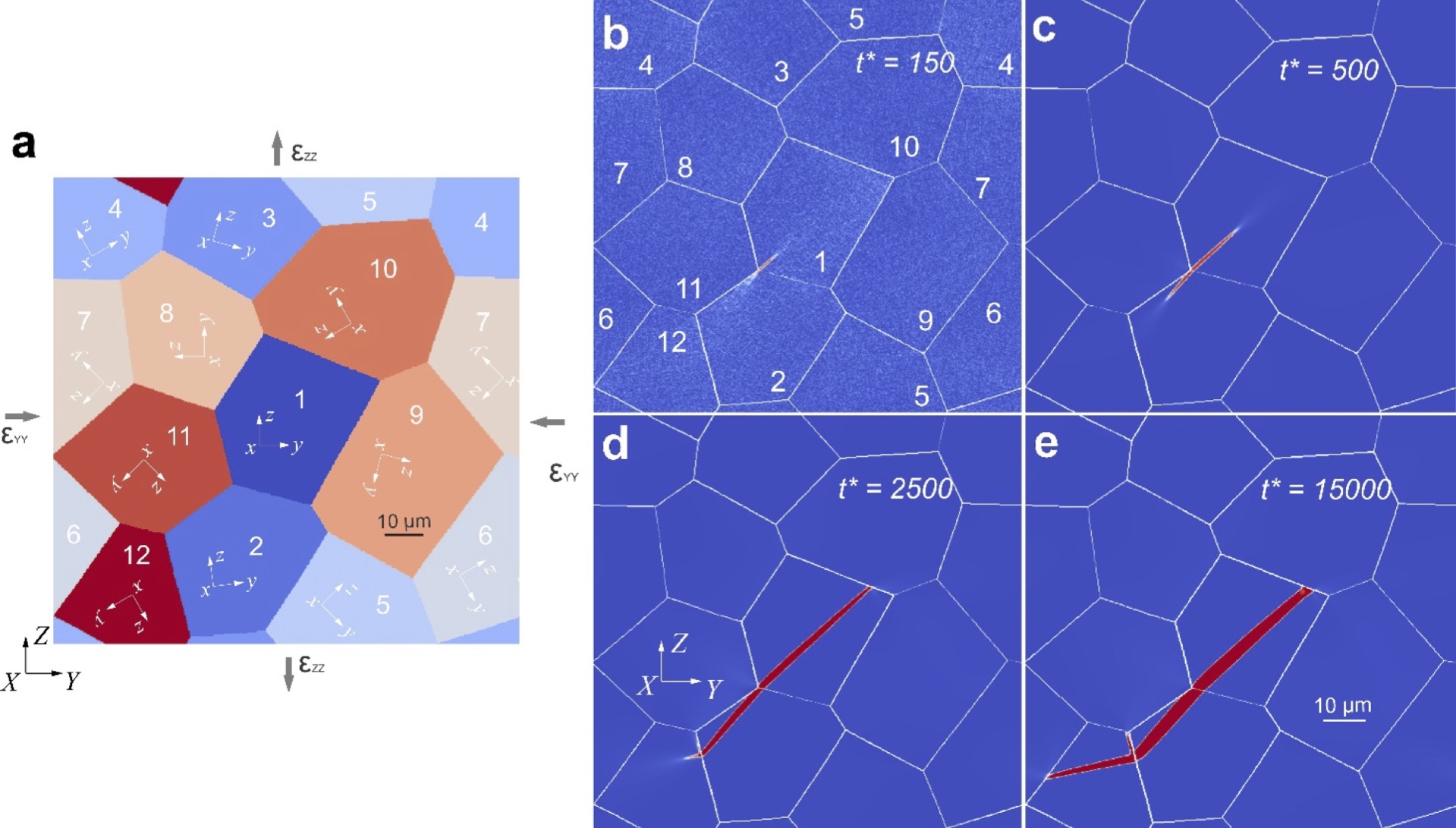}
\caption{
\label{fig:twinning} 
Phase-field simulation of the formation and autocatalytic nucleation of $\{10\bar12\}$ deformation twins in Mg \cite{liu2018formation}: (a) Polycrystalline sample, grain orientations, and applied strain along Y and Z directions, and (b-e) time evolution of the nucleation and growth of twins under an applied strain with $\varepsilon_{YY} = -0.0005$ and $\varepsilon_{YY} = 0.0005$.
}
\end{figure}

A recent application of the PF method \cite{liu2018formation} to the study of the formation and autocatalytic nucleation of $\{10\bar12\}$ twins in a polycrystalline Mg microstructure is depicted in Fig.~\ref{fig:twinning}. Under the effect of an applied strain (Fig.~\ref{fig:twinning}a), a twin crystal nucleates at a triple-junction (Fig.~\ref{fig:twinning}b), grows and triggers the formation of other twin crystals in the neighboring grains, which ultimately form a twin chain (Fig.~\ref{fig:twinning}c,d) that thickens until a steady state microstructure is reached (Fig.~\ref{fig:twinning}e). In particular, this study has shed light on the links existing between twin nucleation and residual shear stresses, stress-free transformation strain of the emerging twins, grain orientations, and stress concentrated at grain boundaries

\section{Phase-Field modeling of Solidification}

\subsection{Fundamental concepts}

Solidification has been a particularly successful application field of the PF method.
Models for solidification mostly rely on the approach proposed by Langer \cite{langer1987chance}, which considers a diffuse interface at an arbitrary scale below that of the microstructure of interest. 
The phase field $\phi$ is thus a global measure of order between solid and liquid phases \cite{boettinger2002phase}. 
By using a unique phase field, rather than several order parameters, the intrinsic crystalline anisotropy is not naturally rendered, but anisotropic interfacial properties (e.g. excess free energy and kinetic coefficient) can be introduced {\it ad hoc} through an orientation dependence of the gradient energy coefficients \cite{kobayashi1994numerical, mcfadden1993phase, karma1996phase, boettinger2002phase, bottger2006phase, haxhimali2006orientation, dantzig2013dendritic}.

An important feature of these models is the existence of well-defined free-boundary equations, which have analytical solutions in both sharp and diffuse interface formulations \cite{steinbach2009phase}. 
On the one hand, the existence of these solutions imposes severe constraints on the model formulation --- for instance, it may be challenging to ensure that the driving force for the phase transformation be constant across the interface \cite{steinbach2009phase, steinbach2013phase, plapp2016phase}. 
On the other hand, spurious effects arising from the use of an arbitrarily wide interface can be quantitatively corrected, since the deviation from a known solution can be quantified.
Quantitative formulations have thus been developed for ``mesoscopic'' interface widths, e.g. using matched asymptotic analysis \cite{karma1996phase, karma1998quantitative, echebarria2004quantitative} and solute ``antitrapping'' terms \cite{karma2001phase, echebarria2004quantitative}, which allow increasing the interface width for numerical convenience, as long as it remains small enough to accurately resolve the interface curvature \cite{caginalp1986phase, caginalp1986higher, steinbach2009phase}.

PF models for solidification are built on fundamental concepts of thermodynamics \cite{boettinger2002phase, steinbach2009phase, plapp2016phase}. 
General formulations may use an entropy \cite{bi1998phase} or a grand-potential \cite{plapp2011unified, choudhury2012grand} functional, but the majority of models are based on the construction of a free energy functional \cite{moelans2008introduction, plapp2016phase}. 
The thermodynamic foundations of such models allow (1) a natural integration of some fundamental physics, such as interface curvature (Gibbs-Thomson) effects and (2) a clear correlation of all model parameters to physical properties, e.g. mobility terms $M_\phi$ and $M_c$ naturally relating to the interface kinetic coefficient and solute diffusion coefficient, respectively.

\subsection{Highlighted applications}

Details and equations of PF models for solidification were already reviewed in details in several articles \cite{ode2001recent, boettinger2002phase, karma2003phase, singer2008phase, steinbach2009phase, provatas2011phase, steinbach2013phase, plapp2016phase}. 
Here, we focus on a few recent examples of application of the method to quantitatively simulate and explain the formation of complex microstructures from the liquid state.

Early PF simulations of solidification (e.g. \cite{kobayashi1994numerical}) were qualitative by nature, in spite of striking morphological similarities between computed patterns and actual microstructures. 
Simulations became more quantitative around the turn of the century with the development of quantitative models for ``upscaled'' interface widths \cite{karma1996phase, karma1998quantitative, karma2001phase, echebarria2004quantitative}, which opened the way to three-dimensional simulations at experimentally relevant length and time scales, with the support of  high-performance computing 
\cite{george2002parallel, nestler20053d, vondrous2014parallel, yamanaka2011gpu, shimokawabe2011peta, provatas1998efficient, provatas1999adaptive, greenwood2018quantitative,  lan2003adaptive, rosam2008adaptive, bollada2015three, guo2012phase}.

A major success of the PF method has been to provide a quantitative simulation tool to discuss classical theories and address long-standing questions in the field of dendritic growth. 
PF simulations have, for instance, permitted to validate microscopic solvability theory \cite{langer1980instabilities, ben1984pattern, benamar1993theory, brener1993needle, kurz2019progress}, and to investigate the morphology of three-dimensional dendritic tips \cite{karma2000three, plapp2000multiscale}.  

Progressively, PF results have been compared to experimental measurements on a more quantitative basis.  Comparisons were made on microstructural length scales measured on solidified samples, e.g. scaling laws for secondary dendrite arm spacings in cast Al-Cu ingots \cite{ode2001numerical, bower1966measurements}, or from {\it in situ} imaging of solidification experiments on transparent organic systems \cite{greenwood2004crossover, liu1995thin, kirkaldy1995thin} or more recently using X-ray imaging of metallic alloys \cite{clarke2017microstructure, boukellal2018scaling}.
Looking at kinetic effects, the dependence of dendrite growth velocity with undercooling was predicted by a quantitative model \cite{bragard2002linking} using MD-calculated parameters for interface energy \cite{hoyt2001method, hoyt2003atomistic} and kinetic coefficient \cite{hoyt1999kinetic} and exhibited a remarkable agreement with experimental data measured in highly undercooled Ni \cite{lum1996high, willnecker1989evidence}. 

The examples above represent a small --- and by no means exhaustive --- sample of efforts to quantitatively compare PF results with solidification experiments and theories. 
Below, we highlight in further details two outstanding applications of PF simulations that have lead to quantitative predictions and explanations of experimentally observed phenomena in solidification. 

\subsubsection{Growth orientation transitions in metallic alloys}

\begin{figure}[!b]
\centering
\includegraphics[width=\columnwidth]{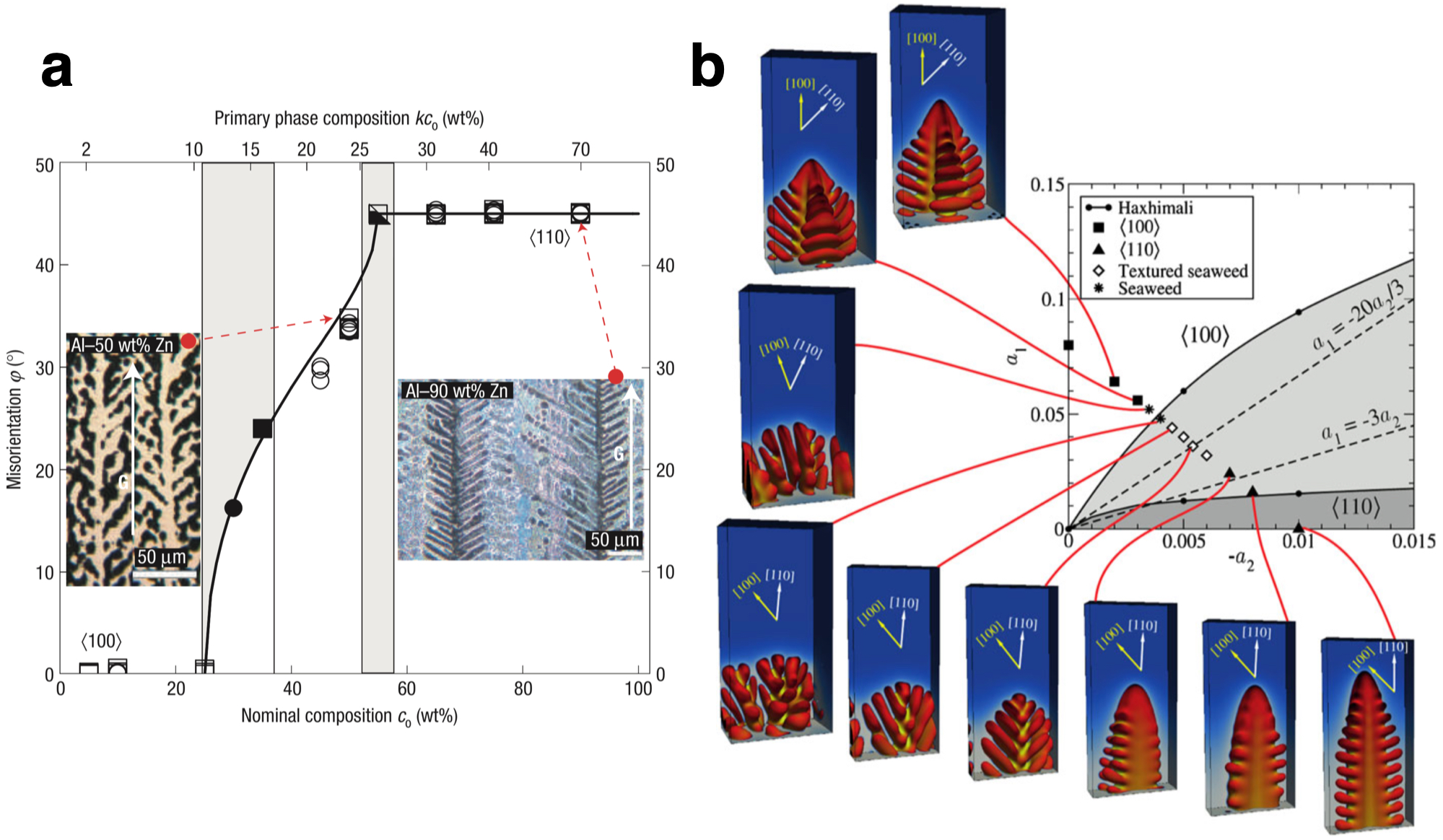}
\caption{
\label{fig:morph_trans} 
Transition of preferred dendritic growth direction in Al-Zn alloys: (a) experimentally observed continuous transitions from $\langle100\rangle$ to $\langle110\rangle$ growth direction as a function of Zn content \cite{haxhimali2006orientation} and (b) corresponding phase-field simulation showing the influence of solid-liquid free energy anisotropy parameters \cite{dantzig2013dendritic}.
Reproduced with permission from (a) \cite{haxhimali2006orientation} and (b) \cite{dantzig2013dendritic}. 
}
\end{figure}

Unconventional ``feathery’’ dendrites have been reported for decades \cite{herenguel1948procedes}. More recently, on the basis of electron backscatter diffraction (EBSD) analysis, it was suggested that a change of solid-liquid free energy due to solute addition might be at the origin of morphological transitions in Al alloys \cite{henry1998dendrite, semoroz2001ebsd}. This assumption was validated for Al-Zn alloys by a combination of solidification experiments and quantitative phase-field simulations \cite{gonzales2006dendrite, haxhimali2006orientation} (Fig.~\ref{fig:morph_trans}). 
Solidification experiments had revealed a continuous transition from $\langle100\rangle$ to $\langle110\rangle$ preferred dendritic growth orientation as the alloy Zn content was increased \cite{gonzales2006dendrite, haxhimali2006orientation} (Fig.~\ref{fig:morph_trans}a). 
The transition was consistent across different setups and different cooling rates, thus pointing at an effect of interfacial energy rather than kinetics. 
This interpretation was validated by quantitative PF simulations \cite{haxhimali2006orientation}. After extending the expression of the solid-liquid interface free energy anisotropy, PF simulations revealed that small changes in the anisotropy parameters were sufficient to trigger this orientation transition in both equiaxed \cite{haxhimali2006orientation} and columnar \cite{dantzig2013dendritic} (Fig.~\ref{fig:morph_trans}b) growth. Independently, molecular dynamics (MD) simulations on a model Lennard-Jones system revealed that alloying may give rise to sufficient changes in interface free energy anisotropy \cite{becker2007atomistic}.

In the past few years, several other dendrite orientation transitions (DOT) were reported, mostly in Al and Mg alloys --- two elements with weak solid-liquid interfacial anisotropy \cite{sun2006crystal, morris2002complete, napolitano2002experimental}. In Al-Ge, a transition was observed as a function of the Ge content, by combining X-ray {\it in situ} imaging, EBSD analysis, and PF modeling \cite{becker2019dendrite}. 
Another recent study has discussed the underlying mechanisms of a DOT in Al-Sm alloys, combining experiments, phase-field, and atomistic simulations \cite{wang2020controlling}. The authors suggest that the mismatch of lattice constants between Al and Sm (rather than liquid ordering) is the reason for the transition, and they report unexpected dependence of interface energy and its anisotropy upon Sm content and temperature.
Other examples of observed orientation transitions include several Mg alloys \cite{casari2016alpha, yang2016effect, du2017correlation, shuai2018synchrotron}, as well as another type of transition from a $\langle100\rangle$ to a $\langle111\rangle$ growth direction in Al-Cu droplets as a function of their cooling rate \cite{bedel2015characterization}.

\subsubsection{Solidification experiments in microgravity}

Classical theories of solidification have focused on the interplay of interface capillarity with diffusive transport of heat and/or species \cite{ivantsov1947temperature, benamar1993theory, kurz2019progress}. However, in the vast majority of solidification processes, the release of heat and solute at the solid-liquid interface induces density inhomogeneities in the liquid that, under the effect of gravity, lead to buoyant convection. This phenomenon, acknowledged for decades \cite{mehrabian1970interdendritic, thi1989influence, dupouy1989natural}, has prevented the solidification of bulk samples in homogeneous conditions \cite{jamgotchian2001localized}. Hence, most of solidification theory was discussed in light of thin-sample quasi-2D experiments, in which convection is limited (however often not completely suppressed \cite{bogno2011analysis, clarke2017microstructure}) by confinement. These limitations have provided a strong motivation to develop solidification experiments in platforms offering lower gravity levels --- e.g. space shuttles \cite{glicksman1994dendritic, thi2005directional}, parabolic flights \cite{nguyen2017interest}, or the International Space Station (ISS) \cite{marcout2006iac, bergeon2011dynamics}. Hence, after decades of sustained efforts from both modeling and experimental communities, solidification simulations and experiments of bulk samples under homogeneous conditions can now be compared on a quantitative basis at the scale of entire three-dimensional cellular or dendritic arrays.

\begin{figure}[!b]
\centering
\includegraphics[width=\columnwidth]{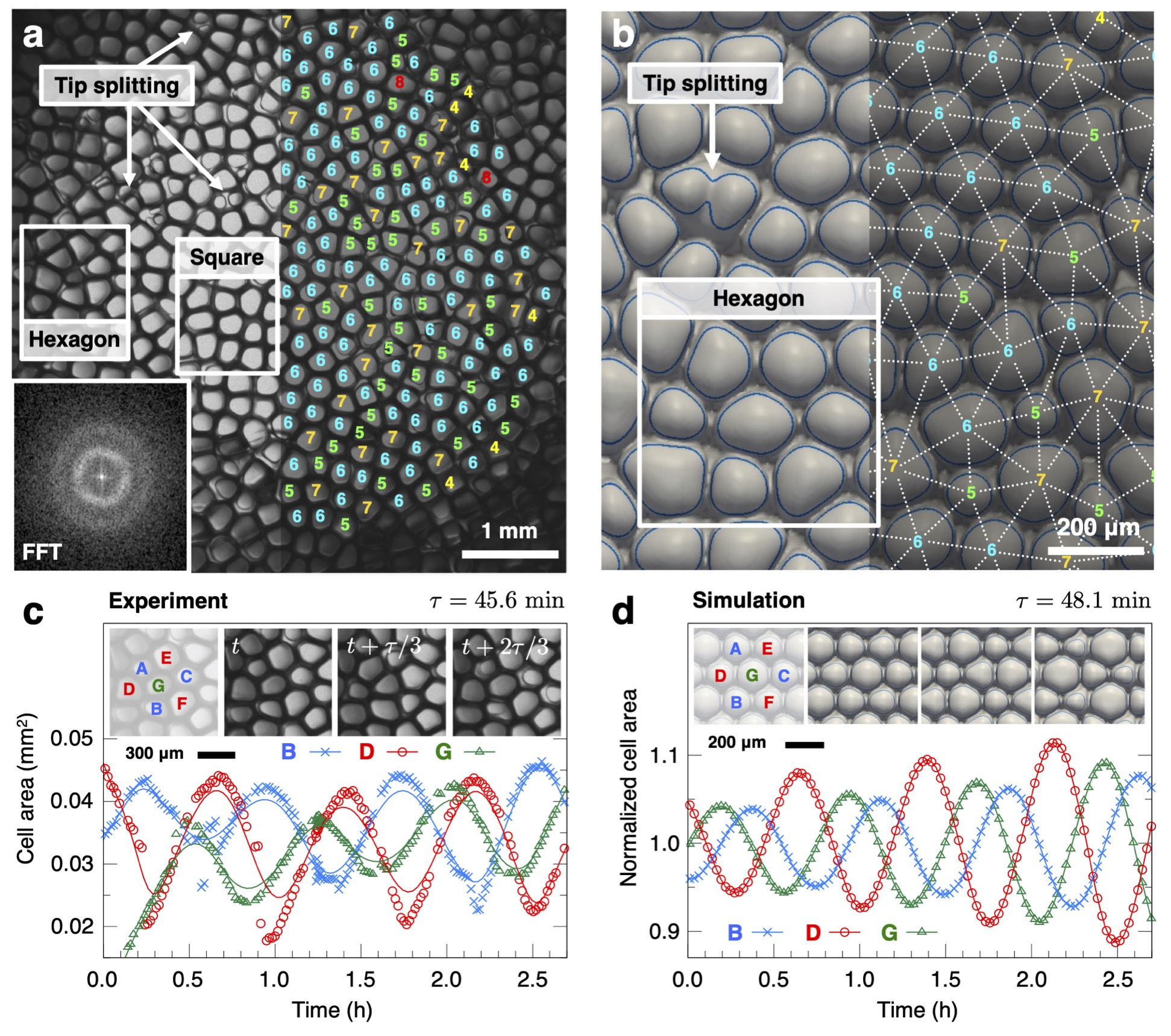}
\caption{
\label{fig:declic} 
Oscillations in three-dimensional cellular patterns \cite{bergeon2013spatiotemporal, tourret2015oscillatory}: (a) experimental {\it in situ} observation and (b) phase-field simulation of the solid-liquid interface, seen from the liquid side facing the growth direction, with tagged numbers of neighboring cells. Locally ordered regions of the array exhibit coherent breathing-mode oscillations. In hexagonal ordered regions, both experiment (c) and simulation (d) exhibit three sublattices oscillating $2\pi/3$ out-of-phase with one another at a period $\tau\approx45$ min.
}
\end{figure}

A recent example of study combining microgravity solidification with phase-field simulations focused on experiments performed in the DEvice for the study of Critical LIquids and Crystallization (DECLIC) installed aboard the ISS \cite{marcout2006iac, bergeon2011dynamics}. This setup allows directional solidification of transparent organic alloys (widely used as model systems for metals \cite{jackson1965transparent, kurz2019progress}), with {\it in situ} imaging of the solid-liquid interface pattern formation. Experiments on dilute succinonitrile-camphor alloys revealed an unexpected oscillatory behavior of cellular arrays, which was explained using phase-field simulations (Fig.~\ref{fig:declic}) \cite{bergeon2013spatiotemporal, tourret2015oscillatory, pereda2017experimental}. 
For several experiments within a narrow range of control parameters (namely temperature gradient and growth velocity), oscillations of the individual cell areas were observed across entire arrays, with an oscillation period between about 25 and 125 minutes. 
The oscillations did not exhibit the long-range spatial coherence that had been observed in thin-sample experiments performed on Earth \cite{georgelin1997oscillatory}. 
However, small regions with local spatial order showed coherent oscillations of subgroups of cells. Quantitative phase-field simulations (Fig.~\ref{fig:declic}b) were capable of reproducing the overall array behavior (Fig.~\ref{fig:declic}a) with a remarkable agreement on the oscillation period (Fig.~\ref{fig:declic}c,d). By allowing to scan a broad range of parameters, simulations allowed to identify the link between this oscillatory behavior and a gap in the array stable spacing range, occurring within a narrow range of temperature gradients. The lack of overall oscillation coherency was attributed to frequent tip-splitting events (Fig.~\ref{fig:declic}a-b) promoting the global disorder of the array. 
Later quantitative analyses of DECLIC experiments with quantitative phase-field simulations have explored the evolution of the three-dimensional cell tip shape (reconstructed through interferometry) during this oscillatory regime \cite{pereda2020experimental}, the effect of sample size and grain orientation on the appearance of the oscillations \cite{mota2020oscillatory}, and the importance of accounting for heat transport kinetics in order to quantitatively predict initial transient dynamics of the planar interface destabilization and cell/dendrite spacing selection \cite{mota2015initial, song2018thermal}. 
Ongoing and future investigations along that research line shall shed light on the formation and stability of three-dimensional grain boundaries \cite{mota2021effect}, as well as the effect of the array ordering on microstructure selection in cellular and dendritic arrays \cite{strickland2020nature, bellon2021multiscale}. 

\section{Perspectives and Challenges}

The examples presented above demonstrate that PF models are at the core of future developments within the framework of Integrated Computational Materials Engineering (ICME) \cite{olson1997computational, NAE08, allison2011integrated}, whose goal is to integrate available modeling tools into a multiscale strategy capable of linking processing, structure, properties, and performance of engineering materials \cite{BXL19}. By providing a consistent thermodynamic framework with a natural integration of interfaces and kinetics at the scale of microstructures, PF occupies a central role into this ICME paradigm. This leads to a number of novel approaches and challenges for future developments, some of which are briefly outlined below.

\subsection{Identification of parameters for quantitative simulations}

The energy functional at the core of PF models depends on a large number of parameters that control the different energy contributions (e.g. chemical, interface, elastic deformation) as well as the atomic and interfacial mobilities in Cahn-Hilliard and Allen-Cahn equations. Quantitative predictions of the microstructure evolution requires a careful identification of parameters and constitutive relations, from experiments when possible, but often from other simulation methods. 

A natural link exists between PF and CalPhaD methods, since both rely on a free energy description, and the two methods have been coupled successfully in various applications \cite{zhu2002linking, steinbach2007calphad, fries2009upgrading}. 
However, the CalPhaD method depends upon databases of experimental, theoretical, and/or computational data, which may be missing for new alloys, or inexistent for interfacial properties, hence requiring atomistic and/or first principles simulations \cite{liu2007integration}. 

For instance, cluster expansion strategies in combination with statistical mechanics can be used to predict the phase diagrams and the Gibbs free energies of different phases (including metastable phases) for alloys \citep{ZJ03, NSP16, LML20}. The interface energies between matrix and precipitates for different crystallographic orientations can be computed using density functional theory \citep{RV18, EMP19, liu2017multiscale, ji2014predicting, liu2013simulation, liu2017structure, liu2017formation, vaithyanathan2004multiscale, ji2018phase, dewitt2017misfit}, and grain boundary energy and mobility can be determined by means of MD simulations \cite{ratanaphan2015grain, rohrer2011grain, rollett2004grain}. For solidification, MD simulations can also be used to determine the solid-liquid interface energy \cite{morris2002complete, hoyt2003atomistic, sun2004crystal, sun2006crystal, zhou2013calculation, asadi2015two, asadi2015quantitative, asadi2016anisotropy, kavousi2019modified} as well as its kinetic coefficient \cite{hoyt1999kinetic, hoyt2002atomistic, sun2004crystal, monk2010determination, mendelev2010molecular, gao2010molecular, kavousi2019combined, kavousi2020interface}. 
However, for such simulations to be predictive, one must pay special attention to selecting interatomic potentials that remain accurate up to the melting temperature (see e.g. \cite{asadi2015quantitative, asadi2015two, asadi2016anisotropy, kavousi2019modified}).

\subsection{Computational cost and acceleration strategies}

\subsubsection{Algorithms and parallelization}

An important limitation of the PF method is its computational cost, which restricts the length and time scales of applications. This limitation can be addressed, to some extent, by numerical techniques (e.g. parallelization~\cite{george2002parallel, nestler20053d, vondrous2014parallel, yamanaka2011gpu, shimokawabe2011peta}, 
adaptive meshing~\cite{provatas1998efficient, provatas1999adaptive, greenwood2018quantitative, lan2003adaptive},
explicit time stepping~\cite{bollada2015three, guo2012phase, rosam2008adaptive},
or spectral methods~\cite{chen1998applications, zhu1999coarsening, feng2006spectral}), but the development of new formulations that remain accurate while using wider diffuse interfaces (or coarser numerical grids) is still actively pursued. 

\subsubsection{New model formulations}

Among newly proposed formulations, a remarkable new method was recently proposed that relies on an intrinsically discrete formulation, and remains accurate (rotational invariance, no grid pinning) with coarser spatial discretization than usually required \cite{finel2018sharp}. 
This new method was recently applied to grain growth, taking advantage of its computational advantage to perform simulations on a statistically significant number of grains leading to a scale-invariant regime \cite{dimokrati2020s}.

In solidification, the thin interface limit \cite{karma1996phase, karma1998quantitative, echebarria2004quantitative} has demonstrated its computational benefits, and even as it remains to be performed for the majority of published models, this type of analysis has now been performed for a range of models (e.g. \cite{folch2005quantitative, kim2007phase, choudhury2012grand}). 
Unequal finite phase diffusivities complicates the thin-interface asymptotic matching \cite{almgren1999siam}. For alloys, the phenomenological solute anti-trapping concept has been extended to arbitrary diffusivity ratio \cite{ohno2009quantitative,ohno2016variational, fang2013recovering}. An interesting new formulation was recently proposed that introduces a kinetic cross coupling between non-conserved and conserved fields, due to non-diagonal terms in force-flux Onsager relations \cite{brener2012kinetic, boussinot2014achieving,wang2020modeling}, opening a way to quantitatively simulations of phase transformations with finite diffusivities in all phases.

\subsubsection{Integration with machine learning}

Recent promising examples of code acceleration or innovative use of PF simulations have involved machine learning (ML). 
Materials science applications of ML have seen an exponential growth over the past years \cite{morgan2020opportunities}, and they have an extremely wide range of applications, from text mining to solving partial differential equations (PDEs) \cite{morgan2020opportunities, mueller2016machine, butler2018machine}. 
In this framework, PF simulations can be used as a tool for generating high-throughput synthetic training data \cite{de2021accelerating, shen2019phase,kunwar2020integration}, or as a model in which to plug in machine-learned functions or parameters (e.g. free energies, mobilities) \cite{kunwar2020integration, teichert2019machine,goswami2020transfer}. 
Alternatively, ML can also be used to directly solve PDEs in a fast and approximate manner \cite{raissi2018hidden, samaniego2020energy}.
Recent examples of PF-ML integration include: ML of free energy landscapes to predict the shape and composition of precipitates \cite{teichert2019machine}, PF-trained ML identification of growth rate constants for electromigration in microelectronics solder joints \cite{kunwar2020integration}, or high-throughput PF generation of synthetic microstructure for ML based on two-point correlation and principal component analysis applied to spinodal decomposition \cite{de2021accelerating}.
Further than microstructure evolution, ML was also used in combination with a range of PF models, for instance for fracture mechanics simulations \cite{goswami2020transfer} or to predict the breakdown strength of polymer-based dielectrics \cite{shen2019phase}.
Beyond a direct integration with PF, the potential of ML to improve PF simulations is tremendous, for instance providing alternate routes for predictions of phase diagrams, development of accurate interatomic potentials, or automated multidimensional microstructure analysis \cite{morgan2020opportunities,mueller2016machine}.
Importantly, statistically-based methods provide a convenient tool to analyze uncertainty propagation in ICME model chains (see e.g. \cite{attari2020uncertainty}).
In turn, ML also comes with outstanding challenges, related for instance to the size and uncertainty of the training data, or the selection of appropriate training metrics.

\subsection{Nucleation}

Nucleation is of paramount importance to microstructure selection, but still challenging to address quantitatively and predictively with PF. Mesoscale descriptions of nucleation have to rely on approximations. For instance, random noise can be used that satisfies the fluctuation-dissipation theorem \cite{wang1998field, artemev2001three, shen2008finding, granasy2002nucleation, granasy2002crystal, granasy2007phase, warren2009phase}, but in general unphysically strong noise is required to prompt nucleation events.
Other methods rely upon explicitly seeding supercritical nuclei, either from a statistical distribution of inoculants, or explicitly made to match classical nucleation theory while respecting conservation balances \cite{simmons2000phase, wen2003phase, simmons2004microstructural, jokisaari2016nucleation, liu2019precipitation}. In the context of solidification, some of the challenges related to nucleation in PF models have notably been discussed in Refs \cite{granasy2019phase} and \cite{plapp2011remarks}. 

\subsection{Rapid solidification}

The emergence of fusion-based additive manufacturing of metals has revived the interest in quantitative modeling of rapid solidification, when the solid-liquid interface is far from equilibrium \cite{boettinger1988microstructural, jacobson1994rapid, kurz1994rapid}. 
An important consequence of rapid solidification in alloys is solute trapping, which stems from a jump in the chemical potential across the interface \cite{aziz1994transition, aziz1988continuous, boettinger1989theory, galenko1997local, galenko1999model}. Using a realistic interface width, solute trapping is well predicted by conventional PF simulations through solute gradient terms in the free energy functional \cite{wheeler1993phase, ahmad1998solute}, or alternatively introducing a relaxation condition for the solute flux though the interface \cite{kim1999phase}. However, since the amount of solute trapped depends on the interface thickness, modeling the departure from equilibrium and the resulting solute trapping using an upscaled interface width has remained a challenge. 

In the finite interface dissipation model, separate compositions profiles for solid and liquid phases are linked by a kinetic equation describing the exchange of components between phases \cite{steinbach2012phase, zhang2012phase,zhang2015incorporating}.
The model introduces an effective interface permeability, an important parameter \cite{reuther2019solute}, which can be calibrated, for instance, by matching with calculations at smaller scale with physical interface width \cite{zhang2013diffuse}. 
Alternative methods include using a relaxation equation for the partition coefficient \cite{kim2021phase}, or prescribing interface conditions through thermodynamic extremal principle \cite{wang2013application}.
Recently, a pragmatic approach was also promposed that consists in modifying the conventional anti-trapping current term \cite{karma2001phase}, performing a thin-interface asymptotic analysis on the resulting equations, and calibrating the parameters of the modified anti-trapping term to an analytical rapid solidification model, typically using the classical continuous growth model \cite{pinomaa2019quantitative, kavousi2021quantitative}.

\subsection{Coupling microstructure evolution with micromechanics}

Another important area of development is the application of PF models to simulate the microstructural changes associated with thermo-mechanical processing, such as grain fragmentation and recrystallization taking place by the combination of temperature and mechanical stresses. Coupling of PF models of microstructure evolution with finite element of FFT solvers for the mechanical deformation appears as a most promising route to allow the ``virtual processing'' and ``virtual design'' of novel microstructures with optimized properties for specific applications. This coupling requires a careful analysis of the differences in the time scales associated with the microstructure evolution and the application of the mechanical stresses. Particularly interesting approaches to achieve these goals have been developed by coupling crystal plasticity and PF models \citep{abrivard2012phase1, abrivard2012phase2, kondo2014phase, GBH14, VBS15, LAO16, ADMAL2018, zhao2020finite}.

\subsection{Multiscale strategies and ``upscaling''}

From the mesoscale upwards, the PF method operates at the ideal length scale to simulate microstructural representative volume elements (RVE) for bridging with models at higher length scales. Upscaling strategies include, for instance: direct calculation of physical parameters (e.g. specific interface areas \cite{neumann2017general} and permeability \cite{takaki2019permeability, mitsuyama2020permeability} of dendritic arrays); benchmarking to calibrate numerical parameters for coarse-scale models (e.g. applied to columnar growth competition with a cellular automaton \cite{pineau2018growth}); and RVE sampling and statistical analysis (e.g. applied to the precipitation and growth of $\delta’$ precipitates in Al-Li alloys \cite{schwarze2018computationally}).

\subsection{Computational benchmarks}

The development of standard benchmark problems has emerged as a pressing matter throughout the PF community as PF modeling has become a standard computational engineering tool. This is important in order to not only test computational efficiency, but also guarantee the accuracy and reproducibility of simulations \cite{jokisaari2017benchmark, jokisaari2018phase, jokisaari2020phase, kamachali2018numerical, eiken2018ses}. 

\subsection{Perspectives and novel applications}

The generality and ease of implementation of the phase-field method makes it an ideal tool for simulations of microstructure formation and evolution at the ``mesoscale''.
Its applications go much further than those highlighted in this short review. 
It is expected that the PF method will play an important role into linking length and time scales in the simulation of advanced manufacturing, such as in additive manufacturing processes of metals (e.g. \cite{abdeljawad2019sintering, yang20193d, lu2018phase, ji2018understanding, karayagiz2020finite}), or microstructure evolution in new classes of materials (e.g. metallic foams \cite{vakili2020controlling,vakili2020multi}).
The PF method also provides an outstanding tool to predict and control the development of micro-/nano-patterned materials, such as nano-templated eutectics \cite{kulkarni2020archimedean} or ice-templated structures \cite{karma2021tms} for instance used in biomedical applications \cite{yin2019freeze} or manufacturing of magnetic composites \cite{yin2021superior}. 
Furthermore, phase-field simulations can help understanding the formation of complex hierarchical biomaterials with outstanding properties \cite{wegst2015bioinspired} (e.g. mollusk shells \cite{schoeppler2019crystal, sun2021crystal}).
In conclusion, in spite of intrinsic scale limitations and remaining fundamental challenges, phase-field models should most certainly play a central role in future discovery and design of innovative materials with outstanding properties for both structural and functional applications.

\section*{Acknowledgements}
This investigation was supported by the European Research Council under the European Union's Horizon 2020 research and innovation programme (Advanced Grant VIRMETAL, grant agreement No. 669141) and by the HexaGB project of the Spanish Ministry of Science (reference RTI2018-098245).
D.T. acknowledges support from the European Union's Horizon 2020 research and innovation programme through a Marie Sk\l odowska-Curie Individual Fellowship (Grant Agreement 842795).
Valuable discussions with Prof. N. Moelans are gratefully acknowledged.


\end{document}